\documentclass[useAMS,usenatbib]{mn2e}
\usepackage{amssymb,amsmath} 
\usepackage[dvips]{graphicx}
\usepackage{fixltx2e}
\usepackage{url}

\include{journals}

\newcommand{\numax}{\mbox{$\nu_{\rm max}$}}

\newcommand{\muHz}{\mbox{$\mu$Hz}}
\newcommand{\kep}{\mbox{\textit{Kepler}}}
\newcommand{\totalsample}{168}
\newcommand{\notblends}{81}
\newcommand{\blends}{87}
\newcommand{\confirmedblends}{18}
\newcommand{\blendsminus}{69}

\title[Kepler red giants with close companions]
{Evidence for compact binary systems around \kep\ red giants}

\author[I. Colman et al.]{Isabel L. Colman$^{1, 2}$\thanks{icol6407@uni.sydney.edu.au}, Daniel Huber$^{1, 2, 3, 4}$, Timothy R. Bedding$^{1, 2}$, James S. Kuszlewicz$^{2, 5}$, \newauthor Jie Yu$^{1, 2}$, Paul G. Beck$^{6}$, Yvonne Elsworth$^{2, 5}$, Rafael A. Garc\'{i}a$^{6}$, Steven D. Kawaler$^{7}$, \newauthor Savita Mathur$^{8}$, Dennis Stello$^{1, 2, 9}$, and Timothy R. White$^{2}$\\
$^{1}$Sydney Institute for Astronomy, School of Physics, A28, University of Sydney, NSW, 2006, Australia\\
$^{2}$Stellar Astrophysics Centre, Department of Physics and Astronomy, Aarhus University, Ny Munkegade 120, DK-8000 Aarhus C, Denmark\\
$^{3}$Institute for Astronomy, University of Hawai`i, 2680 Woodlawn Drive, Honolulu, HI 96822, USA\\
$^{4}$SETI Institute, 189 Bernardo Avenue, Mountain View, CA 94043, USA\\
$^{5}$School of Physics and Astronomy, University of Birmingham, Edgbaston, Birmingham B15 2TT, UK\\
$^{6}$Laboratoire AIM, CEA/DRF - CNRS - Univ. Paris Diderot - IRFU/SAp, Centre de Saclay, 91191 Gif-sur-Yvette Cedex, France\\
$^{7}$Department of Physics and Astronomy, Iowa State University, Ames, IA 50011, USA\\
$^{8}$Center for Extrasolar Planetary Systems, Space Science Institute, 4750 Walnut street Suite 205, Boulder, CO 80301, USA\\
$^{9}$School of Physics, University of New South Wales, NSW 2052, Australia
\\}

\begin{document}

\date{Accepted --. Received --; in original form --}


\maketitle

\label{firstpage}

\begin{abstract}
We present an analysis of \totalsample\ oscillating red giants from NASA's \kep\ mission that exhibit anomalous peaks in their Fourier amplitude spectra. These peaks result from ellipsoidal variations which are indicative of binary star systems, at frequencies such that the orbit of any stellar companion would be within the convective envelope of the red giant. Alternatively, the observed phenomenon may be due to a close binary orbiting a red giant in a triple system, or chance alignments of foreground or background binary systems contaminating the target pixel aperture. We identify \blends\ stars in the sample as chance alignments using a combination of pixel Fourier analysis and difference imaging. We find that in the remaining \notblends\ cases the anomalous peaks are indistinguishable from the target star to within 4$''$, suggesting a physical association. We examine a Galaxia model of the \kep\ field of view to estimate background star counts and find that it is highly unlikely that all targets can be explained by chance alignments. From this, we conclude that these stars may comprise a population of physically associated systems.
\end{abstract}


\section{Introduction}
\label{sec:intro}

\begin{figure}
\begin{center}
\includegraphics[width=1\linewidth]{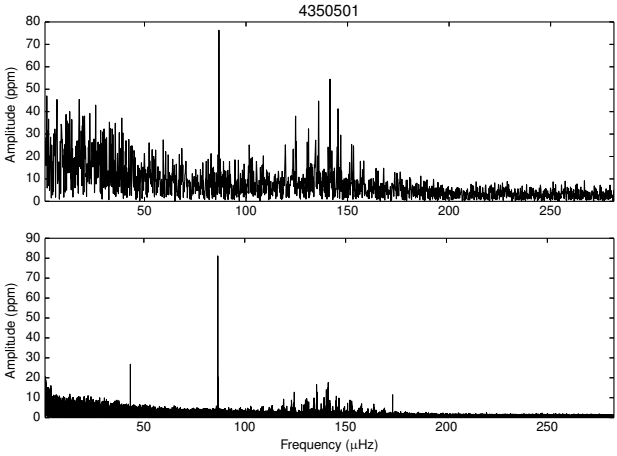}
\caption{Top panel: Amplitude spectrum of KIC~4350501, the first noted red giant with an anomalous peak, using only 43 days of data (Q0 and Q1), as in \protect\cite{bedding_solar-like_2010}. Bottom panel: Amplitude spectrum calculated using all four years of \kep\ data. The oscillations are centred at 140\,\muHz; the anomalous peak is at $\sim$86\,\muHz.}
\label{fig:eg}
\end{center}
\end{figure}

\begin{figure*}
\begin{center}
\includegraphics[width=1\linewidth]{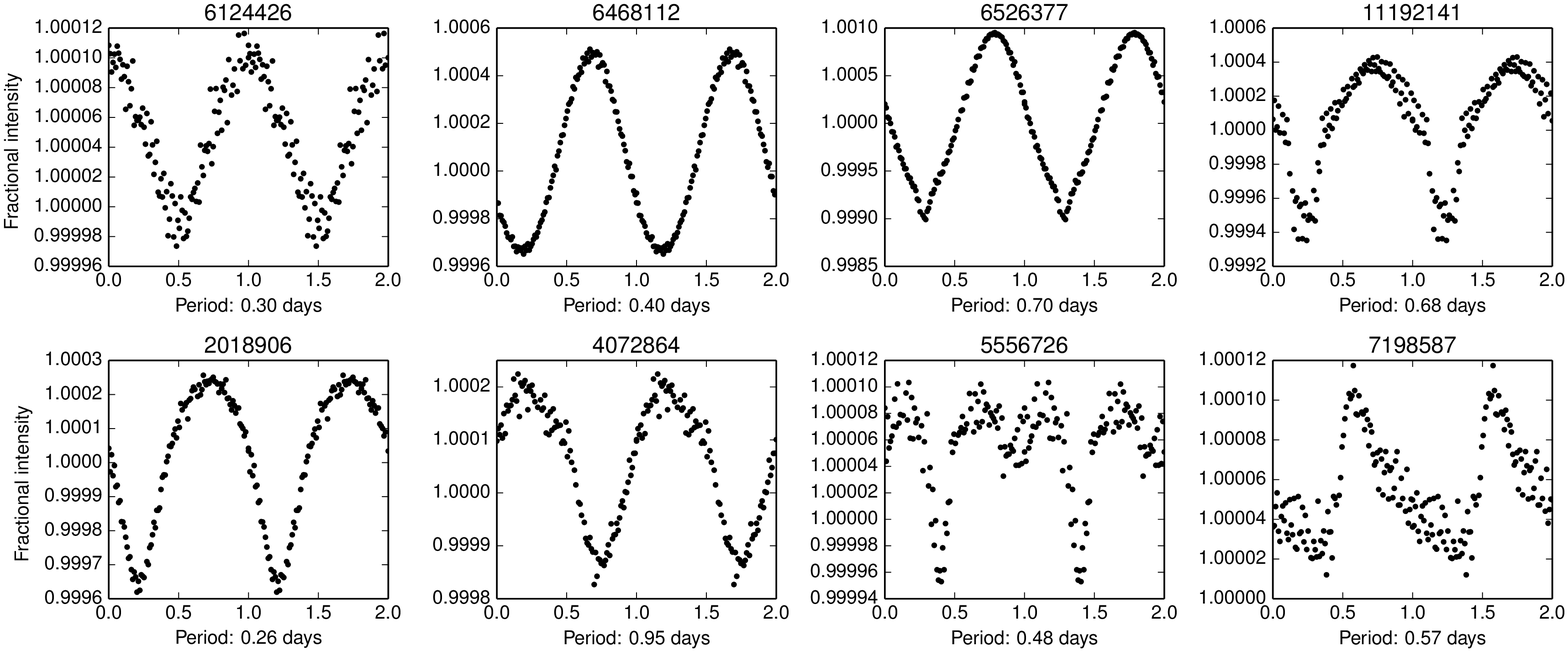}
\caption{A variety of light curves phased on the period of the anomalous peak and binned. The top row shows stars that could not be discounted as chance alignments in this study. The bottom row shows chance alignments.}
\label{fig:phase}
\end{center}
\end{figure*}

This paper sets out to solve a long-standing problem in the study of oscillating \kep\ red giants. The analysis of red giants has been an area of rapid growth with the advent of data from the \kep\ mission~\citep{borucki_kepler_2010}. In particular, asteroseismology has allowed unprecedented insights into their core fusion~\citep{bedding_gravity_2011}, internal rotation~\citep{mosser_spin_2012, beck_fast_2012}, and internal magnetic fields~\citep{stello_prevalence_2016}. Red giants have also contributed to the field of galactic archaeology, where the study of red giant populations is used to map the formation history of the galaxy~\citep{miglio_galactic_2013, casagrande_measuring_2016}.

The red giants in this study were first noted in the early days of \kep. Initial analysis revealed anomalous high-amplitude peaks in their Fourier spectra. Figure~\ref{fig:eg} shows amplitude spectra using the first quarter of data and data from all quarters for the first star observed to exhibit this behaviour, KIC~4350501~\citep{bedding_solar-like_2010}. Note that the heights of the oscillation modes in the amplitude spectrum decrease as the observing time is increased because the modes become more resolved.

These anomalous peaks were first suggested to be mixed modes~\citep{bedding_solar-like_2010}, which are caused by the coupling between p modes propagating in the convective envelope with g modes propagating in the radiative core. Solar-like oscillations are stochastically excited and damped, with narrower peaks indicating longer mode lifetimes, as expected for mixed modes. However, this peak does not conform to the typical comb-like pattern of red giant oscillations~\citep{hekker_giant_2016}. In the case of KIC~4350501, more data showed the anomalous peak to be intrinsically narrow and revealed a subharmonic at half the frequency of the peak, which implies that it is a distinct signal, unrelated to the red giant oscillations. This suggests that the peak is not a mixed mode but rather the signature of tidal interactions in a binary system. We have subsequently found many other red giants that show this type of behaviour, including the presence of harmonics and subharmonics. However, these peaks are present at such short periods that any binary companion would have to be orbiting within the convective envelope of the red giant.

This raises the possibility that we are observing common-envelope systems~\citep{paczynski_common_1976}. This is a phase of binary evolution that has been extensively studied with modelling and population synthesis. Evidence to confirm the existence of common-envelope systems is hard to come by; the closest method we have to direct detection is studying observational phenomena indicative of a past common-envelope phase. Recent studies have used the shaping of planetary nebulae with binary central stars to better understand common-envelope interaction~\citep{hillwig_observational_2016}, and jets in planetary nebulae to constrain the magnetic fields of common-envelope binaries~\citep{tocknell_constraints_2014}. It has also been postulated that the common-envelope phase could be integral to the evolution of red giants into sdB stars and cataclysmic variables~\citep{beck_pulsating_2014}. The observation of a common-envelope system would provide important confirmation for these theories of binary evolution. For a recent review of our understanding of common-envelope systems, see \cite{ivanova_common_2013}.

Another possibility is that these objects may be examples of hierarchical triple systems, where a compact binary orbits a red giant, e.g. HD181086 (``Trinity")~\citep{derekas_hd_2011, fuller_tidally_2013}. Alternatively, these anomalous peaks could arise from a chance alignment: a background or foreground compact binary that has contaminated the light collected from the red giant. This study examines a sample of \totalsample\ light curves that exhibit both red giant oscillations and an anomalous peak, often with harmonics or a subharmonic. In this paper, we outline the method used to identify chance alignments, and comment on the statistics of possible physically associated systems.

\section{Methods and Analysis}
\label{sec:methods}

\subsection{Data preparation}
\label{sec:data}

\begin{figure}
\begin{center}
\includegraphics[width=1\linewidth]{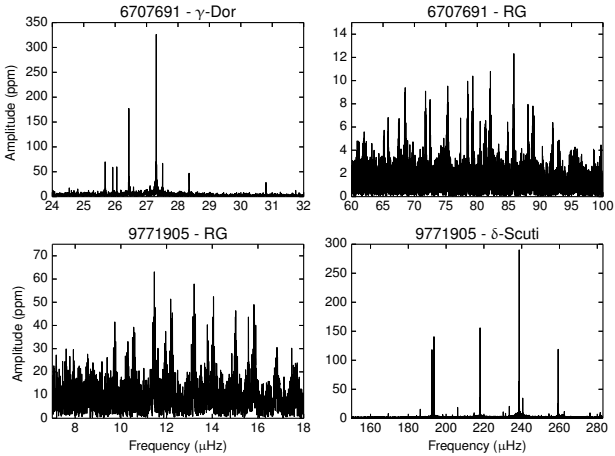}
\caption{Top row: KIC~6707691, showing both $\gamma$-Dor (left) and red giant (right) oscillations. Bottom row: KIC~9771905, showing both red giant (left) and $\delta$-Scuti (right) oscillations. Each set of oscillations is isolated to better display its features, as in each star the red giant oscillations have significantly lower amplitudes than the classical pulsator oscillations.}
\label{fig:eg2}
\end{center}
\end{figure}

The \totalsample\ stars studied were all discovered among \kep\ red giants by visual inspection of power spectra. Many were included in the samples of \cite{huber_asteroseismology_2010}, \cite{huber_testing_2011}, or \cite{stello_asteroseismic_2013}. Additional stars were taken from \cite{yu_asteroseismology_2016} or found in the APOKASC sample~\citep{pinsonneault_apokasc_2014}.

We began by downloading and preparing \kep\ simple aperture photometry (SAP) light curves from MAST\footnote{\url{http://archive.stsci.edu/kepler/}}. We processed the light curves following \cite{garcia_preparation_2011}, initially performing a high-pass filter using a Gaussian of width 100 days. We followed this by clipping all outliers further than 3$\sigma$ from the mean. Finally, we took a Fourier transform to produce the amplitude spectrum.

We first located the comb-like pattern of solar-like oscillations, typical of red giant stars. Then, we were able to identify anomalous peaks. These have no particular position in relation to the solar-like oscillations. The majority of anomalous peaks had amplitudes higher than or comparable to the oscillations. Some anomalous peaks were found at similar frequencies to the oscillations, which led to a degree of confusion in stars that were identified previously with fewer quarters of data. A subset of the stars that we initially considered to fit this pattern were discarded from this study due to the anomalous peak representing an oscillatory $\ell=0$ or $\ell=2$ mode with a relatively broad peak, implying a shorter mode lifetime.

Using the frequency of the high-amplitude anomalous peak, we phase-folded each star's time series. The majority of the resulting phase curves displayed ellipsoidal variation, lending weight to the theory that these peaks are due to binarity. None of the anomalous peaks included in this study displayed the phase variation expected of a red giant oscillation. Examples are given in Figure~\ref{fig:phase}. In some cases there were also subharmonics present in the Fourier spectra, as with KIC~4350501 (Figure~\ref{fig:eg}), or a series of peaks indicative of an eclipse.

Anomalous peaks in seven cases came from nearby main sequence pulsators---two with $\gamma$-Doradus oscillations and five with nearby $\delta$-Scuti pulsators. Examples of red giants contaminated by $\gamma$-Dor and $\delta$-Scuti oscillations are shown in Figure~\ref{fig:eg2}. Although these stars do not conform to the typical pattern of red giant oscillations with one anomalous peak and possible harmonics and subharmonics, we include them in this paper as they were studied with the same processes as the remainder of the sample and provide confirmation that the method works independently of the type of target being analysed.

\subsection{Pixel power spectrum analysis}
\label{sec:pixel}

\begin{figure*}
\begin{center}
\includegraphics[width=1\linewidth]{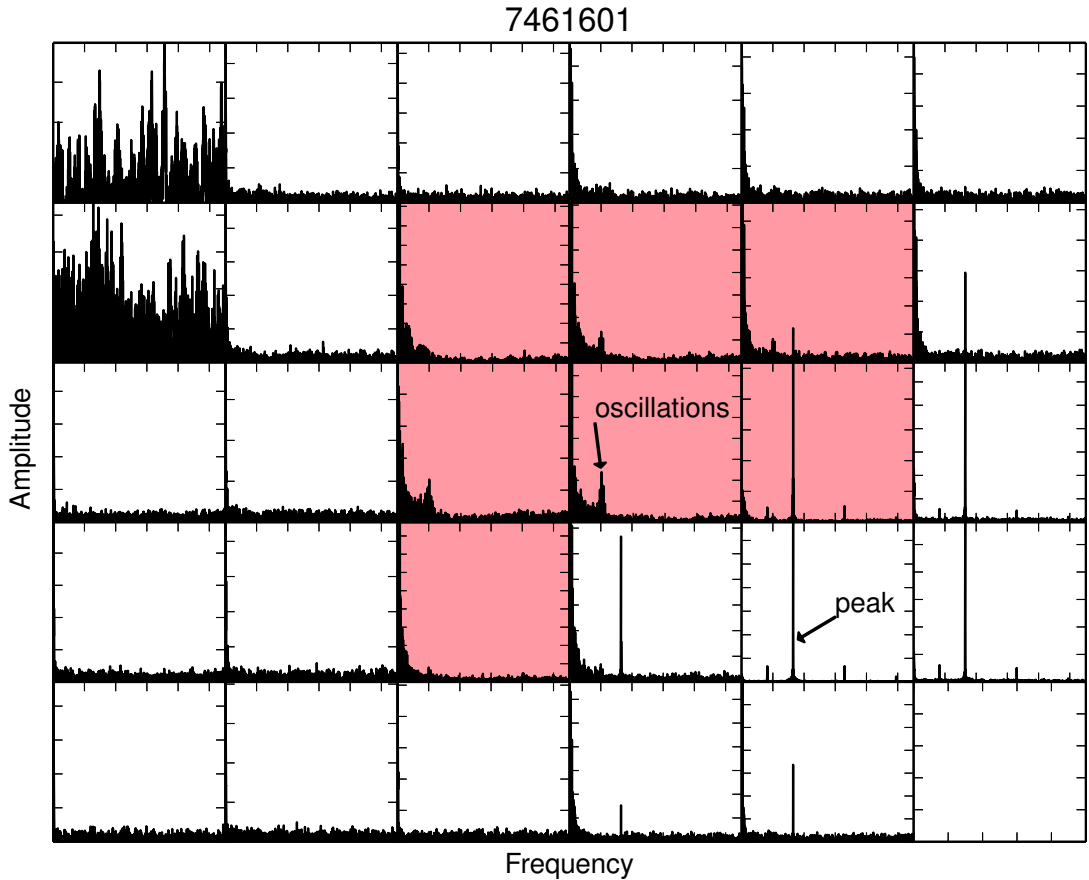}
\caption{The \kep\ TPF aperture of KIC~7461601, a red giant showing contamination from a chance alignment with a binary. Each panel represents a pixel, showing an amplitude spectrum calculated using the same methods as in Figure~\ref{fig:eg} with frequencies up to the \kep\ long cadence Nyquist frequency, 283.21\muHz. Amplitudes in each pixel are auto-scaled in order to better display qualitative features. More compact tick marks indicate higher overall amplitudes. Shaded pixels indicate the optimal aperture.}
\label{fig:pixels}
\end{center}
\end{figure*}

\begin{figure*}
\begin{center}
\includegraphics[width=1\linewidth]{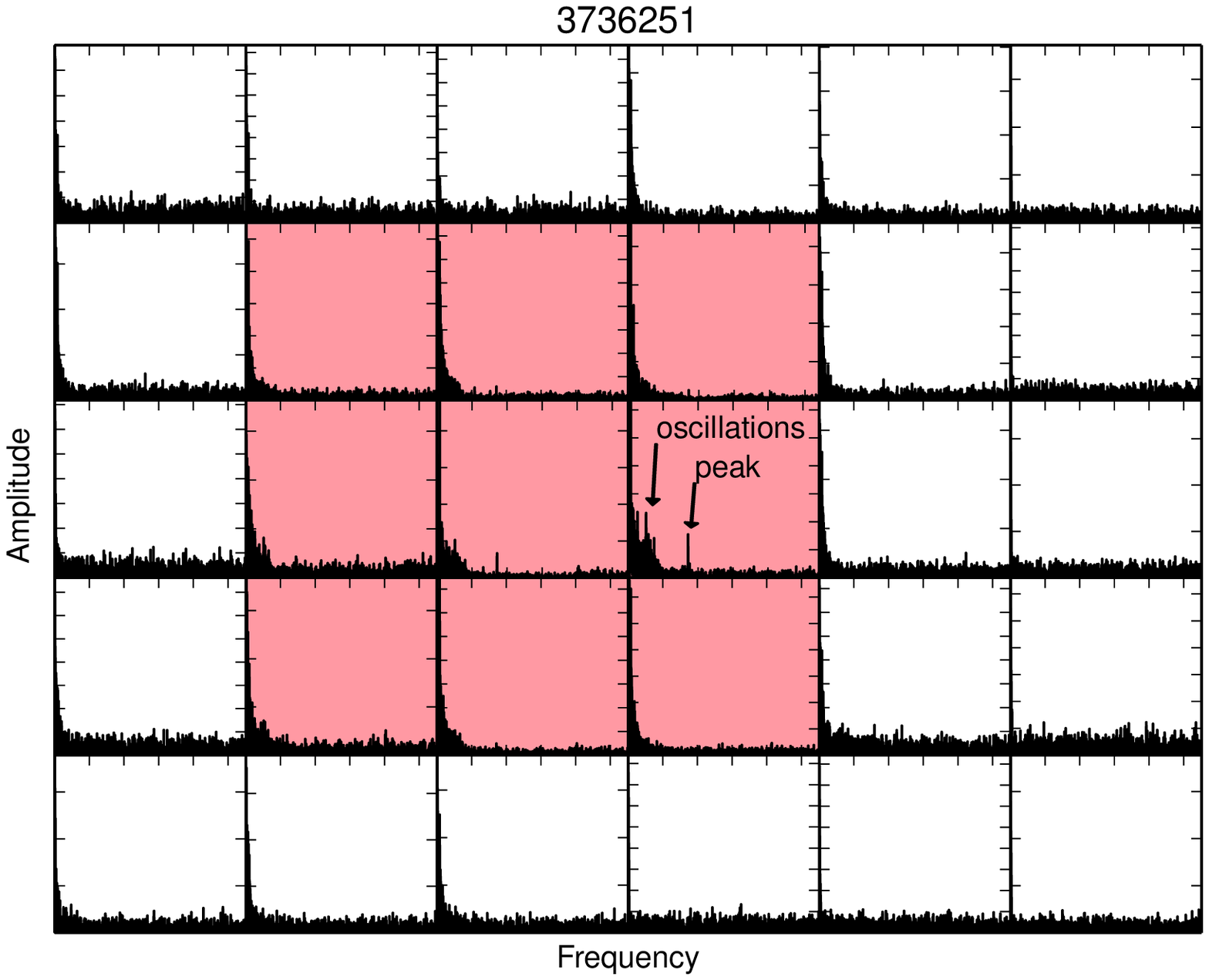}
\caption{The \kep\ TPF aperture of KIC~3736251, showing no contamination. Each panel represents a pixel, as in Figure~\ref{fig:pixels}.}
\label{fig:pixels2}
\end{center}
\end{figure*}

In the previous section we covered the process of identifying stars for this sample. For this, we used SAP light curves, which are a composite of several 4$''$ \kep\ pixels comprising the so-called `optimal aperture.' To locate the true sources of these anomalous peaks, it was necessary to examine the area around each of the targets. For this, we used \kep\ target pixel files (TPFs), which are available for download from MAST. TPFs provide a `postage stamp' image of pixels around \kep\ targets. We employed the same methods outlined in Section~\ref{sec:data} to process light curves from individual pixels in each TPF.

During the \kep\ mission, the orientation of the telescope changed by 90$^\circ$ every quarter. Because of this, we examined each quarter of pixel data separately. In cases where oscillations were not visible with only one quarter of data, we stitched together the light curves of quarters with the same orientation, which occurred every fourth quarter.

By taking a Fourier transform of each pixel time series, we could more accurately locate the source of the anomalous peaks in the image. We identified these by inspection, based on the pixels included in the optimal aperture around the target star. In many cases, the source of the anomalous peak was obviously separated from the source of the solar-like oscillations. Figure~\ref{fig:pixels} shows an example of such a TPF for KIC~7461601, where the optimal aperture is indicated by red shading. In this case, the anomalous peak is primarily located outside the optimal aperture. It is evident that its source is separate to the source of the oscillations. Of the \totalsample\ stars analysed, we found \blends\ to display this type of clear separation. We interpret these as chance alignments of red giants with background or foreground binaries. The other \notblends\ stars did not show this sort of clear separation. Figure~\ref{fig:pixels2} shows the TPF for KIC~3736251, a case where there is no clear distinction between the pixel source of the red giant oscillations and the anomalous peak. We interpret these as possibly physically associated systems.

\subsection{Difference imaging}
\label{sec:image}

\begin{figure}
\begin{center}
\includegraphics[width=0.9\linewidth]{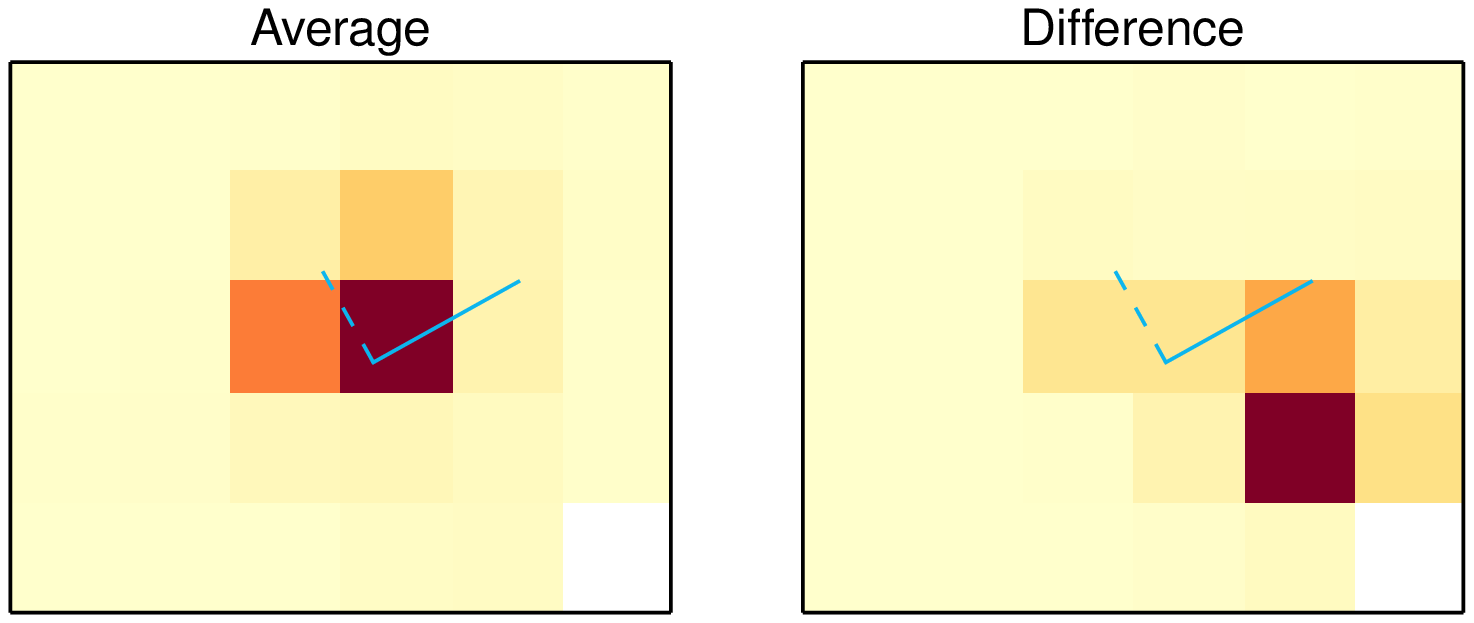}
\caption{An example of difference imaging, displaying the KIC~7461601 aperture as in Figure~\ref{fig:pixels}. To indicate scale, the compass arms are 6$''$. The solid line points north, and the dashed line points east. The variation originating from the bottom right corner of the TPF can be clearly seen in the difference image.}
\label{fig:diff}
\end{center}
\end{figure}

\begin{figure}
\begin{center}
\includegraphics[width=1\linewidth]{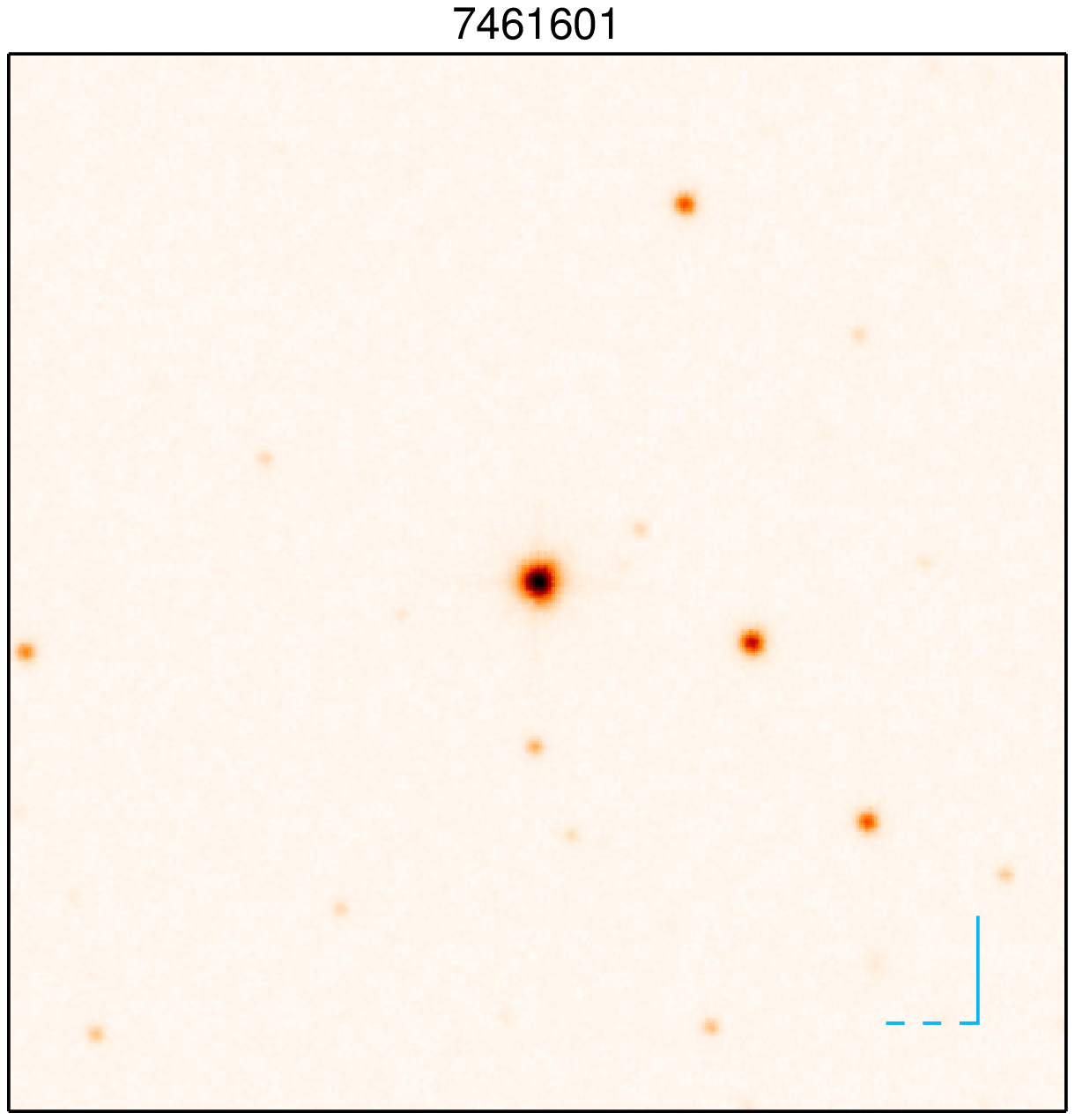}
\caption{A UKIRT image showing a 1$'$ field of view around KIC~7461601. To indicate scale, the compass arms are 6$''$, as in Figure~\ref{fig:diff}. The solid line points north, and the dashed line points east.}
\label{fig:img}
\end{center}
\end{figure}

We performed a more detailed study of the TPFs with difference imaging, which has been successfully applied to the identification of false positive exoplanet transits~\citep{bryson_identification_2013}. We selected postage stamp images that fell in time within 10\% bands centred on the maximum and minimum points of the phased light curve. To create the difference image, we took the average of both sets of images and subtracted the average about the minima from the average about the maxima. This new image retained the dimensions of a TPF postage stamp and could easily be compared to the average images, as shown in Figure~\ref{fig:diff}. From this, we could see which pixels were the source of the flux variations at the period of the anomalous peak.

There remain some caveats for the use of difference imaging. The method we used was designed for ellipsoidal variation, and so it was less useful for the few stars in the sample where the phased light curve showed an eclipse, or where the identified anomalous peak belonged to $\delta$-Scuti oscillations with multiple high-amplitude peaks. There were several other issues with using the phased light curves, particularly in stars with a low signal-to-noise ratio where the periodicity was hard to discern by looking at the phased light curve, due to scatter. Additionally, difference imaging was less successful for cases where the contaminant was at an angular distance greater than 30$''$ from the target star. Some stars with clear contamination in the TPF did not show any variation in the difference image, which suggested that the contaminant was located outside the optimal aperture. This tended to coincide with low-amplitude anomalous peaks. In such cases, it was clear simply from the TPFs that there was contamination.

Despite this, we were still able to gain valuable information from difference imaging. For stars with no evident contamination in the TPFs, the difference images tended not to show variation when compared to the average images. Difference imaging also helped to confirm the status of stars with low signal in the TPF Fourier spectra. Conversely, the difference images reinforced the status of stars with more tentative classification as spatially separated. It follows that many of the cases where difference imaging did not confirm contamination correspond to widely separated chance alignments.

To identify the true source of chance alignments, widely-separated or otherwise, we next compared both the average and difference images to higher-resolution images of the same area of sky, using 1$'$ cutouts from the UKIRT WFCAM (the UK Infrared Telescope Wide Field Camera) survey~\citep{lawrence_ukirt_2007}. An example is shown in Figure~\ref{fig:img}. \kep\ TPFs contain world coordinate system (WCS) information, which allowed us to calculate the orientation of the postage stamp. We displayed coordinates on both types of images in the form of a compass rose, from which we could see whether there were any possible contaminant stars from the same position as the anomalous source as shown in the TPF Fourier spectra. Looking for matches in both the KIC and the UKIRT object catalogue, we were able to use a \kep\ light curve to confirm the source of contamination in \confirmedblends\ cases (see Table~\ref{tab:sample2} in the appendix.) For \blends\ of the other chance alignments (Table~\ref{tab:sample3}), we noted one or more possible stars that could be the contaminant, especially closer to the Galactic plane, which is where most of the chance alignments were found.

\section{Discussion}
\label{sec:results}

\subsection{Spatial distribution}
\label{sec:rgs}

\begin{figure*}
\begin{center}
\includegraphics[width=1\textwidth]{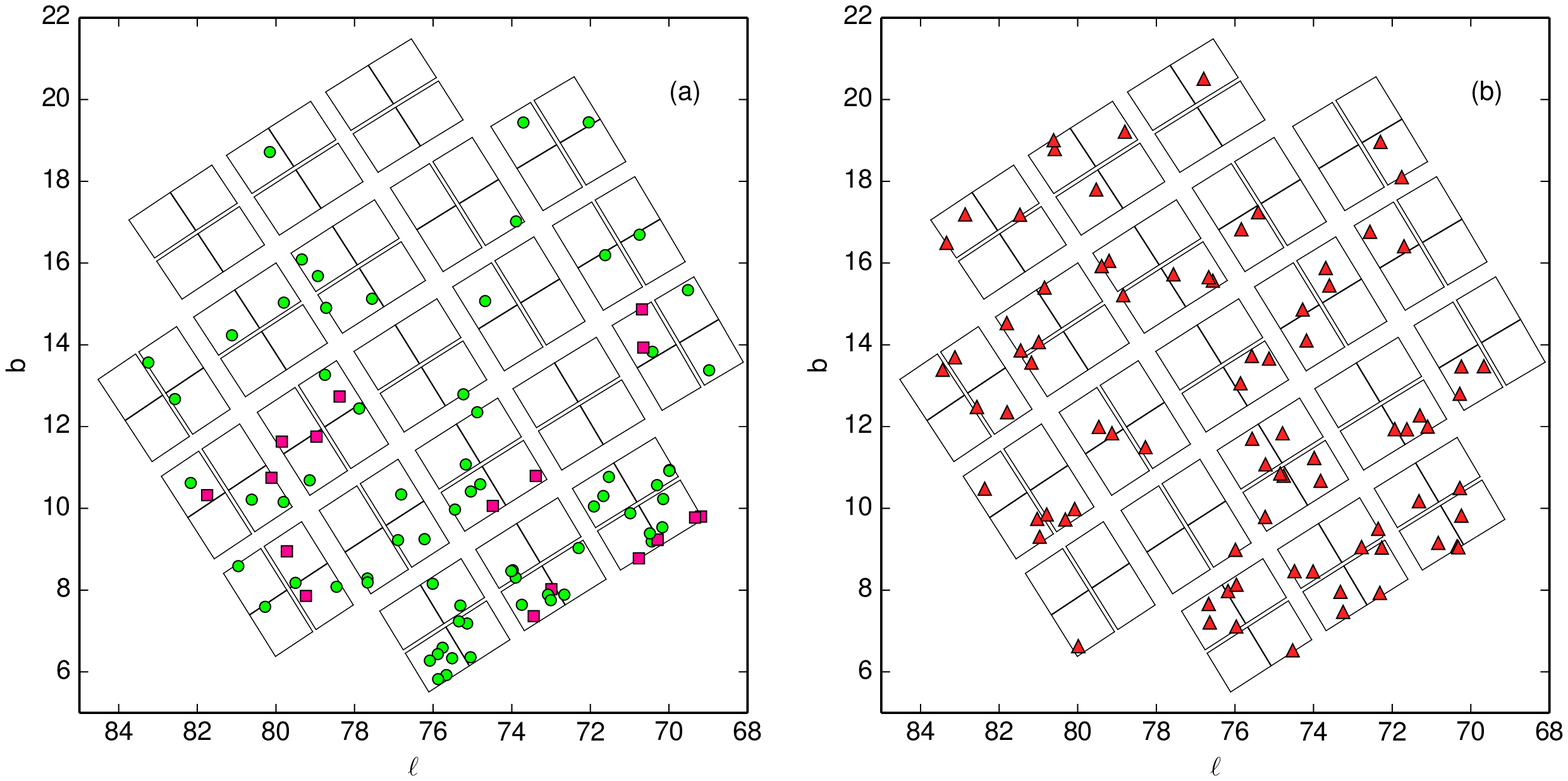}
\caption{The sample of stars in this study is shown across the \kep\ field of view in galactic coordinates. Panel (a) shows red giants with anomalous peaks that we classified as chance alignments, shown as green circles. In cases where source of the contamination could be confirmed by the contaminating star having a \kep\ light curve, the target is shown by a pink square. Panel (b) shows the population of red giants exhibiting an anomalous peak and where a physical separation cannot be discerned.}
\label{fig:populations}
\end{center}
\end{figure*}

\begin{figure}
\begin{center}
\includegraphics[width=0.5\textwidth]{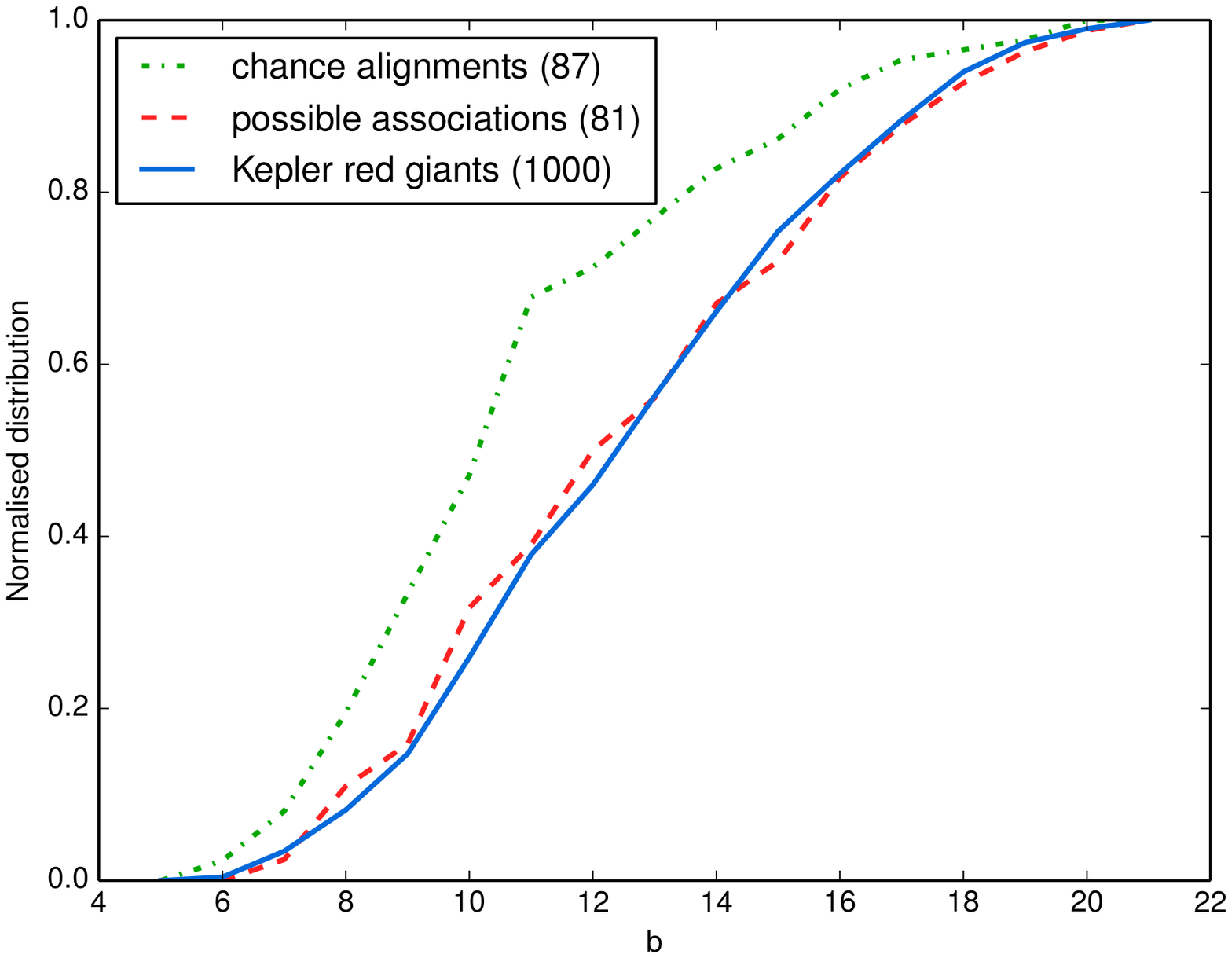}
\caption{Cumulative distributions in galactic latitude of the populations shown in panels (a) and (b) of Figure~\ref{fig:populations}. The solid blue line is taken from a random sample of 1,000 \kep\ red giants.}
\label{fig:cumdist}
\end{center}
\end{figure}

We found that, of the \totalsample\ red giants with anomalous peaks, \blends\ could be identified as chance alignments. For \confirmedblends\ of these chance alignments, we confirmed their status with the analysis of \kep\ light curves of nearby stars which we identified as the sources of contamination. We could not spatially resolve the \notblends\ other stars, and we refer to these as possibly associated systems. We identified four of the five $\delta$-Scuti influenced systems as chance alignments. The two $\gamma$-Dor influenced systems remain possible physical associations.

Figure~\ref{fig:populations} shows the distribution of our sample over the \kep\ field of view (FOV), with chance alignments in panel (a) and possibly associated systems in panel (b). We present these populations in galactic coordinates, and note that the bottom of the field at lower galactic latitudes is closer to the Galactic plane and has a higher density of stars. At higher galactic latitudes we observe a marked paucity of stars as expected, both in the field itself and in the sample considered in this study. Similarly, this pattern presents itself in the distribution of chance alignments. It is therefore noteworthy that the possibly associated systems seen in panel (b) seem to be spread quite evenly across the FOV.

We further analysed these populations by examining their cumulative distributions as a function of galactic latitude, shown in Figure~\ref{fig:cumdist}. We compare this to a distribution of 1,000 red giants drawn randomly from a list of oscillating \kep\ red giants provided from Yu et al. (in preparation). The distribution of the possible physical associations closely matches the distribution of random red giants, which implies that they are not chance alignments. It is also noteworthy that these distributions visibly differ from the distribution of chance alignments, which increases sharply at low galactic latitudes, reflecting the higher density of stars closer to the Galactic plane. The probability of finding a chance alignment between two populations is expected to scale as the square of surface density. This gives us an important insight into the nature of this population and suggests that we may be observing a distinct population of systems, possibly hierarchical triples or common-envelope binaries.

\subsection{Amplitude distribution}
\label{sec:amps}

\begin{figure*}
\begin{center}
\includegraphics[width=1\textwidth]{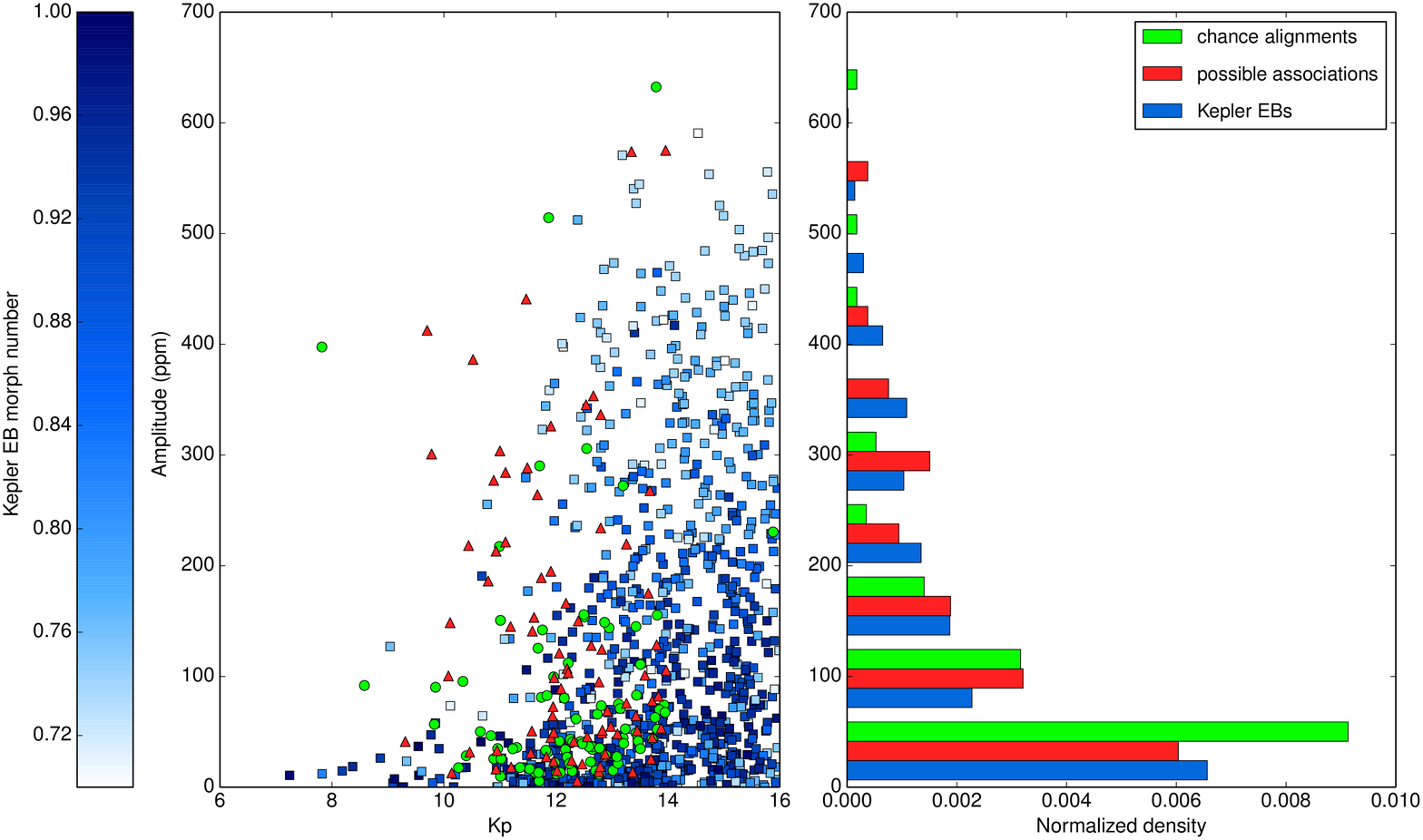}
\caption{Left: Relationship between the amplitude of the anomalous peaks in our sample and \kep\ magnitude. Chance alignments are shown with green circles, and possible physical associations with red triangles. For comparison, we also display the amplitudes of a population selected from the \kep\ Eclipsing Binary Catalog with morphology number $>$ 0.7, shown with blue squares. Right: Histogram of each population distributed across amplitude. Not pictured: two outliers with amplitudes above 700ppm, one chance alignment (KIC~4071950) and one possible physical association (KIC~6526377).}
\label{fig:kp}
\end{center}
\end{figure*}

We searched for a possible correlation between the intrinsic luminosities of the red giants and the amplitudes of their anomalous peaks. It might be expected that if a compact binary is physically associated with a red giant, the amplitude of variations from the binary might correlate inversely with the luminosity of the red giant, due to dilution. We observed no correlation, which led us to compare our stars to a sample drawn from the \kep\ Eclipsing Binary Catalog~\citep{prsa_kepler_2011}\footnote{\url{http://keplerebs.villanova.edu/}}. We selected the sample of eclipsing binaries (EBs) by their morphology, which is a measure of the ellipticity of their phased light curves calculated by locally linear embedding~\citep{matijevic_kepler_2012}. The cut off for EB selection was a morphology number $>$ 0.7, chosen by visual inspection of stars in the catalog to match those with light curves similar to those in our sample. In Figure~\ref{fig:kp}, we plot the amplitudes of the anomalous peaks in our sample and of the \kep\ EBs against \kep\ magnitude. These data show that more ellipsoidal variation tends to have a lower amplitude of variation, a trend which is also present in our sample. The measure of ellipticity in our data was based on a ranking of the shape of phased light curves and on a different scale to the Catalog's morphology number, so we do not display it in Figure~\ref{fig:kp}.

From this exercise, we can explain the lack of correlation between the intrinsic luminosities of red giants and the amplitudes of their anomalous peaks by the broad range of amplitudes present across ellipsoidal variables, as exhibited by the Catalog sample. We also note that while our sample is overall brighter than the Catalog sample, the distribution of amplitudes is what would be expected for a sample primarily exhibiting ellipsoidal variation. The histogram in the right panel of Figure~\ref{fig:kp} shows that the distribution of the possible physical associations closely matches the distribution of the Catalog EBs. The anomalous peaks of both the possible physical associations and the chance alignments are more present at lower amplitudes, but this is markedly noticeable for the latter. This effect can be explained by the wide angular separation between the target stars and their contaminants, so less of the contaminating light enters the optimal aperture. This leads to systematically lower apertures. In the case of the possible physical associations, this dilution could be caused by a compact binary companion. This strengthens the conclusion that the possible physical associations comprise a distinct population.

\subsection{Modelling of chance alignments}
\label{sec:modelling}

\begin{figure}
\begin{center}
\includegraphics[width=0.5\textwidth]{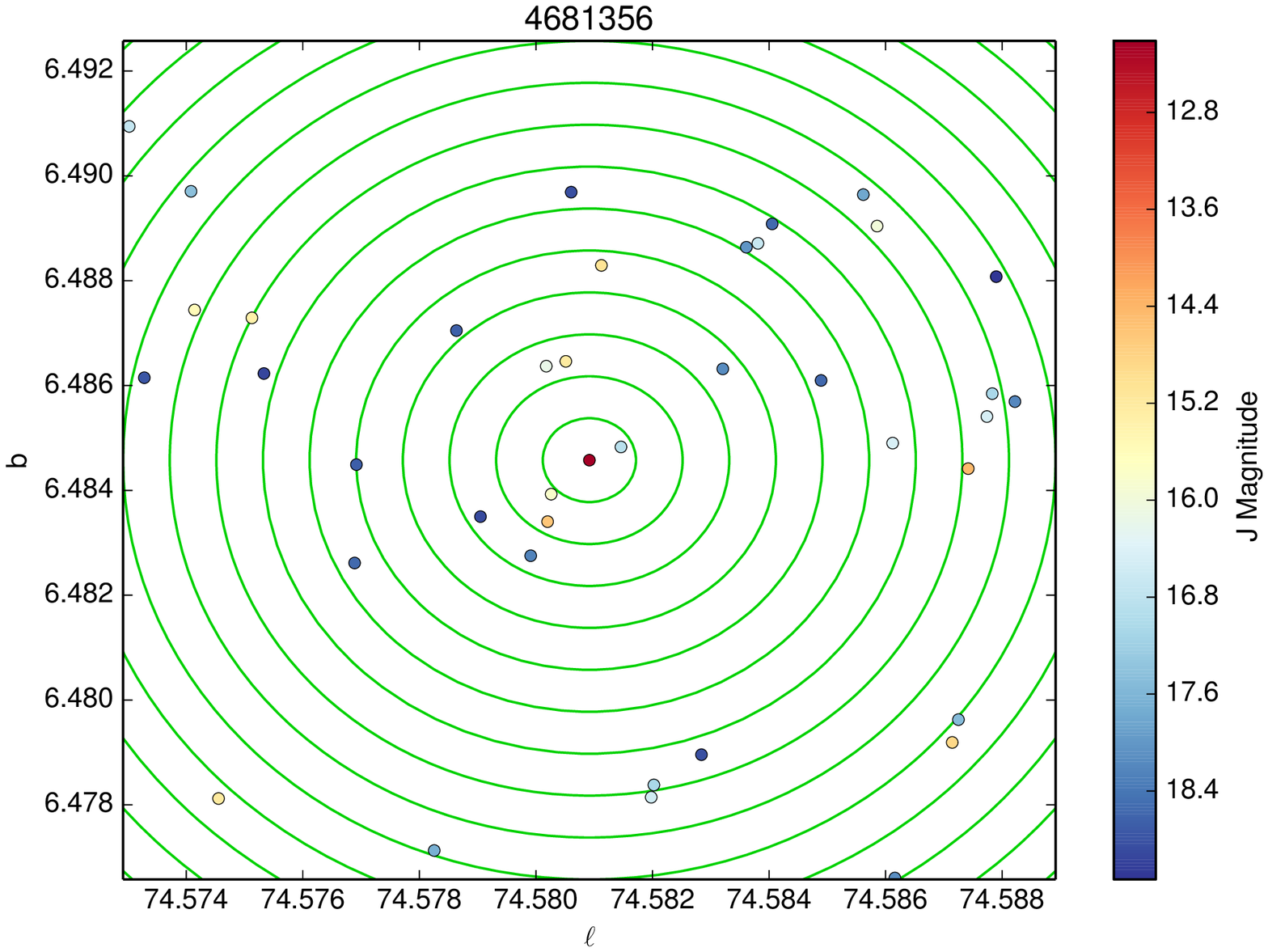}
\caption{An illustration of the process used to search for chance alignments in the Galaxia model, described in Section~\ref{sec:modelling}. The concentric rings have radius 4$''$, intended to represent the maximum distance between stars that could fall on the same \kep\ pixel. The colour scale represents J magnitude.}
\label{fig:model}
\end{center}
\end{figure}

While the majority of chance alignments found in this study involved contaminants further than 4$''$ from the target star, it is possible that there could be contaminants within 4$''$ of the target that our methods do not have the sensitivity to detect. To test whether we could expect to find more chance alignments within the remaining \notblends\ stars, we analysed a model of a stellar population in the \kep\ FOV. This also helped us in understanding the underlying statistics around chance alignments of red giants and background or foreground binary systems.

We used the modelling software Galaxia \citep{sharma_galaxia:_2011}, which allows the user to synthesise an artificial population of stars within a given area of sky, here chosen to match the \kep\ FOV. We defined a chance alignment as any two stars found within a \kep\ pixel of each other, namely 4$''$. Galaxia does not take into account the existence of binary systems, and hence any chance alignments that are detected in the simulation are true chance alignments.

Our model goes down to an apparent $J$ magnitude of 20. Once the synthetic FOV had been simulated, we then searched for stars analogous to those in the sample by minimising over apparent $J$ magnitude within a set range of galactic coordinates $b$ and $\ell$, and stellar parameters $T_{\rm eff}$, log($g$) and [Fe/H]~\citep{mathur_revised_2017}. The nearest match to each target star was designated a blend if we located another star within 4$''$ of it. This process is illustrated in Figure~\ref{fig:model}.

We found the occurrence of 4$''$ chance alignments in the model to be rare, with only 18 of \totalsample\ these stars fitting the criterion, or 10.7\%. This is much lower than the observed fraction (as discussed in Section~\ref{sec:rgs}) because here we are only looking at matches within 4$''$, which cannot be discerned by the techniques covered in Section~\ref{sec:methods}. This can be compared with the figure quoted in a study of false positive KOIs (\kep\ Objects of Interest) by \cite{ziegler_robo-ao_2017}, who found that planet host candidates have a nearby star within 0.15$''$~--~4$''$ with a probability of 12.6\%~$\pm$~0.9. Despite the fact that \citeauthor{ziegler_robo-ao_2017} were not focused on red giants in the same way as our study, our value is just over 2$\sigma$ from the result given by \citeauthor{ziegler_robo-ao_2017}, which places the two samples in good agreement. This suggests that a similar proportion of our sample will contain chance alignments within 4$''$. By inspection of UKIRT images, 9 of the identified chance alignments appear to be close to or within 4$''$. This implies that we should expect to find roughly 9 more chance alignments within 4$''$ among the \notblends\ stars that have not been identified as chance alignments. This is a strong result, and leaves us with a sizeable population of possible physical associations involving an oscillating red giant.

\subsection{Nature of the population}
\label{sec:pop}

Based on the analysis presented in this section, it is likely we are observing a distinct population of possibly physically associated systems. It is evident from the oscillations present that we are observing systems involving red giant stars. This raises the question of how similar the red giants in this sample are to typical red giants. We compared these objects to a sample of $\sim$16,000 known oscillating \kep\ red giants (Yu et al. in preparation) in a plot of \numax\ against oscillation amplitudes, and found that the two samples were similar. This suggests that whatever mechanism is involved in generating large amplitude variations in these systems is not suppressing or altering the red giant oscillations in any significant way.

The most likely possibilities remain common-envelope and hierarchical triple systems. There are no known oscillation common-envelope systems for comparison, and there is still a limit to our knowledge of hierarchical triple systems. The HD181086 system~\citep{derekas_hd_2011} is the best-studied observational example of a hierarchical triple involving a red giant. The red giant in HD181086 does not exhibit any oscillations, and the analyses of red giants in other eclipsing binaries have indicated that red giant oscillations can be suppressed by binarity~\citep{gaulme_surface_2014}. However, since we used the presence of red giant oscillations as a selection criterion for this study it is not possible to observe any such trend. Ultimately, we cannot extrapolate from the case of HD181086 to the many possible cases in this study, so the population of triple systems remains a valid hypothesis.

\section{Conclusions}
\label{sec:conclusions}

From a sample of \totalsample\ red giant stars with anomalous high-amplitude peaks, we found \blends\ could be discounted as chance alignments, with the remaining \notblends\ exhibiting no contamination outside a \kep\ pixel. This leaves the opportunity for these stars to be physically associated systems such as a common-envelope binary or hierarchical triple systems. We observe that this population appears to follow the distribution of randomly-selected stars from the \kep\ FOV. This distinguishes them from the population of chance alignments, which appear with a greater density towards the galactic plane. We have constructed and examined a model of a synthetic population in the \kep\ field which suggests that such close chance alignments are rare, which would imply that most of these stars are more likely to be physically associated systems. This may point to hierarchical triple systems, or to common-envelope binaries.

Future work includes an observation of all remaining targets by Robo-AO~\citep{baranec_robo-ao:_2011}, an adaptive optics system which has been used previously to examine exoplanet host stars and asteroeismic targets observed by \kep~\citep{schonhut-stasik_robo-ao_2017, ziegler_robo-ao_2017}. We will also look to spectroscopic follow-up observations using data from APOGEE~\citep{majewski_apache_2010}, to identify the spectral lines of companion stars or any radial velocity variations indicative of binarity. In addition to this, further opportunities will arise to search for these unusual red giants in data from K2 and TESS. A larger sample from different areas of the sky would aid our understanding of these unusual cases and aid more detailed analysis of a possible new population of systems.

The fundamental parameters of the stars in this study are listed in tables in the appendix. All code used for analysis is available online at \url{https://github.com/astrobel/chancealignments}.

\section*{Acknowledgments}

This paper includes data collected by the \kep\ mission. Funding for the \kep\ mission is provided by the NASA Science Mission directorate. Some of the data presented in this paper were obtained from the Mikulski Archive for Space Telescopes (MAST). STScI is operated by the Association of Universities for Research in Astronomy, Inc., under NASA contract NAS5-26555. Support for MAST for non-HST data is provided by the NASA Office of Space Science via grant NNX09AF08G and by other grants and contracts. The UKIDSS project is defined in \cite{lawrence_ukirt_2007}. UKIDSS uses the UKIRT Wide Field Camera (WFCAM; \cite{casali_ukirt_2007}) and a photometric system described in \cite{hewett_ukirt_2006}. The science archive is described in \cite{hambly_wfcam_2008}. We have used data from the 2nd data release, which is described in detail in \cite{warren_ukirt_2007}. IC acknowledges scholarship support from the University of Sydney. DH acknowledges support by the Australian Research Council's Discovery Projects funding scheme (project number DE140101364) and support by the National Aeronautics and Space Administration under Grant NNX14AB92G issued through the \kep\ Participating Scientist Program. JSK acknowledges the support of the UK Science and Technology Facilities Council (STFC) and the support of the School of Physics and Astronomy, University of Birmingham. PGB, and RAG acknowledge the ANR (Agence Nationale de la Recherche, France) program IDEE (n$^\circ$\,ANR-12-BS05-0008) ``Interaction Des Etoiles et des Exoplanetes". PGB, and RAG also received funding from the CNES grants at CEA. YE acknowledges the support of the UK Science and Technology Facilities Council (STFC). Funding for the Stellar Astrophysics Centre (SAC) is provided by The Danish National Research Foundation (Grant agreement no.: DNRF106). SM would like to acknowledge support from NASA grants NNX12AE17G and NNX15AF13G and NSF grant AST-1411685. DS is the recipient of an Australian Research Council Future Fellowship (project number FT1400147). TRW acknowledges the support of the Villum Foundation (research grant 10118).

\bibliographystyle{mn2e}
\bibliography{rgpaper2}

\clearpage
\appendix

\begin{table*}
\begin{small}
\begin{center}
\caption{Details of the \notblends\ possible physical associations. Stellar parameters are taken from the NASA Exoplanet Archive, data release 25~\citep{mathur_revised_2017}. In cases where we found the anomalous peak to be part of $\gamma$-Dor oscillations, the peak frequency ($\nu_{\rm peak}$) is marked with an asterisk. We mark $\delta$-Scuti anomalous peaks with two asterisks.}
\vspace{0.1cm}
\begin{tabular}{r l l r r r r r r r}
\hline
KIC & RA (deg) & DEC (deg) & Kp & log(g) & $T_{\rm eff}$ (K) & {[}Fe/H{]} & $\nu_{\rm max}$ ($\mu$Hz) & $\nu_{\rm peak}$ ($\mu$Hz) & $P_{\rm peak}$ (days)\\  
\hline
1726211&292.50447&37.29278&10.93&2.39&4981&-0.74&31.56&66.01&0.18\\
1726245&292.51086&37.25521&11.57&2.59&4837&0.21&53.66&43.78&0.26\\
2160901&291.5993895&37.53966&12.06&2.67&4676&0.24&56.54&2.97&3.89\\
2449020&292.68245&37.75239&11.91&2.84&5007&0.07&66.76&13.98&0.83\\
2573092&290.85716&37.87599&11.58&2.46&4723&0.07&31.68&44.69&0.26\\
3356438&294.92994&38.47866&11.97&2.85&4999&0.07&58.40&5.94&1.95\\
3530823&287.101181&38.60348&11.74&2.56&5036&0.07&44.55&21.27&0.54\\
3546046&291.8050695&38.64798&11.96&3.18&4845&0.16&186.12&7.71&1.50\\
3736251&288.20211&38.8702&13.59&3.38&5148&-0.72&25.97&85.71&0.14\\
3858714&293.65017&38.95757&11.94&2.61&4852&0.21&48.27&61.02&0.19\\
3973137&296.026121&39.0659&13.65&2.41&4926&-0.58&36.93&2.56&4.53\\
4043436&287.40072&39.10863&12.77&2.42&4653&0.10&31.36&15.61&0.74\\
4149966&289.5747495&39.25492&10.08&2.79&4934&0.07&72.19&5.02&2.30\\
4164236&293.17628&39.24902&13.97&2.47&4738&-0.04&35.29&61.29&0.19\\
4279165&295.51473&39.36227&12.38&2.61&4868&-0.16&46.14&222.12&0.05\\
4374169&293.93991&39.41256&11.67&2.67&4891&0.07&40.07&7.83&1.48\\
4456739&289.362849&39.54505&12.02&2.44&4658&0.36&41.78&61.63&0.19\\
4555699&289.94031&39.69181&12.80&2.57&4768&-0.08&26.92&2.63&4.40\\
4681356&297.897023&39.709683&13.45&2.57&4710&-0.36&47.16&63.68&0.18\\
4830095&290.10768&39.96584&13.10&2.45&5166&-0.50&30.92&53.11&0.22\\
5112950&295.37307&40.20586&12.77&2.53&4753&0.00&41.51&91.43&0.13\\
5462460&295.65894&40.62042&12.40&2.41&4999&-0.50&32.70&40.99&0.28\\
5793628&292.67057&41.06844&11.10&2.48&4878&-0.48&36.04&49.55&0.23\\
5985252&298.17206&41.23463&11.00&2.31&4936&-0.50&27.47&37.37&0.31\\
6124426&292.1016&41.46219&13.88&3.68&5406&-0.28&205.92&39.16&0.30\\
6185964&284.57787&41.55982&12.98&2.65&4852&-0.42&27.75&39.86&0.29\\
6382801&296.9842395&41.73708&13.72&2.68&4724&0.28&38.10&69.47&0.17\\
6451664&294.57057&41.89605&12.56&2.45&4996&0.07&35.19&50.16&0.23\\
6462755&297.30729&41.84648&10.44&2.53&4785&-0.16&27.53&32.96&0.35\\
6468112&298.50825&41.8638&9.70&3.06&5089&-0.04&64.40&*29.22&0.40\\
6526377&293.07981&41.94772&11.81&2.59&4789&0.00&32.30&16.59&0.70\\
6610354&293.0651805&42.04949&9.31&2.61&4883&-0.20&45.46&7.74&1.50\\
6707691&295.9976805&42.17442&11.91&2.89&5122&0.07&85.82&*27.31&0.42\\
6716840&298.00289&42.11263&11.91&2.55&4841&0.36&27.57&40.50&0.29\\
6753216&282.487031&42.22581&11.49&3.34&5149&-0.74&46.87&**245.20&0.05\\
6929104&284.55087&42.46167&13.80&2.96&5054&-0.22&23.77&41.97&0.28\\
6948654&291.84099&42.43724&13.96&3.08&5048&0.10&32.78&36.40&0.32\\
6952430&293.019909&42.47953&11.83&2.48&4784&0.07&36.22&61.48&0.19\\
7267370&286.715829&42.88117&12.32&2.56&4835&0.07&43.74&61.69&0.19\\
7272332&288.70806&42.86038&13.26&2.63&4779&-0.14&46.89&58.60&0.20\\
7418275&281.62991&43.00333&13.37&3.41&5369&-0.10&221.62&57.55&0.20\\
7447072&292.4606&43.05733&13.26&2.93&5059&-0.16&33.26&23.22&0.50\\
7511777&286.22405&43.12891&13.72&3.53&5140&0.08&221.24&12.05&0.96\\
7596350&287.81166&43.25247&11.10&2.58&5153&-0.50&39.24&43.92&0.26\\
7816294&289.78314&43.52288&11.47&2.62&4644&0.18&48.01&25.45&0.45\\
8092097&289.94732&43.93543&12.80&2.43&4781&-0.20&24.50&33.33&0.35\\
8095225&290.95461&43.9071&13.43&3.32&5268&-0.04&79.96&76.07&0.15\\
8462775&301.53743&44.40842&10.89&2.68&4828&0.02&33.92&52.81&0.22\\
8870432&285.3189&45.1689&9.78&2.53&4733&0.56&35.00&45.39&0.25\\
9008090&286.0825605&45.3942&12.79&2.49&4785&0.07&38.46&67.39&0.17\\
9029195&294.38823&45.33999&10.93&2.52&4848&0.21&40.43&11.68&0.99\\
9146423&288.12324&45.55873&10.95&2.15&4499&-0.02&18.73&54.12&0.21\\
9210116&288.06693&45.68367&10.11&2.75&4912&0.07&53.87&43.65&0.27\\
9541892&297.49509&46.18827&12.67&2.54&4809&0.07&34.31&35.88&0.32\\
9605626&297.981431&46.26839&13.83&3.06&5122&-0.38&33.35&48.30&0.24\\
9763419&288.4587&46.50697&11.55&2.41&4934&0.07&32.20&49.54&0.23\\
9777198&294.5120895&46.59932&12.63&2.48&4775&0.21&36.39&40.76&0.28\\
9851743&298.97724&46.61877&10.46&2.54&4869&0.07&44.81&60.31&0.19\\
9908646&298.160649&46.73501&13.68&2.97&5021&-0.14&23.58&40.08&0.29\\
9969574&298.46844&46.88932&12.09&2.96&4812&0.30&108.63&4.00&2.90\\
\hline
\end{tabular}
\end{center}
\end{small}
\end{table*}

\begin{table*}
\begin{small}
\begin{center}
\begin{tabular}{r l l r r r r r r r}
\hline
KIC & RA (deg) & DEC (deg) & Kp & log(g) & $T_{\rm eff}$ (K) & {[}Fe/H{]} & $\nu_{\rm max}$ ($\mu$Hz) & $\nu_{\rm peak}$ ($\mu$Hz) & $P_{\rm peak}$ (days)\\  
\hline
10334585&289.86435&47.43528&12.82&2.42&5023&-0.50&31.14&51.69&0.22\\
10384595&281.514071&47.50767&12.20&2.87&5186&0.07&44.83&55.09&0.21\\
10724041&288.92067&48.08529&12.28&2.41&4847&0.21&28.08&75.24&0.15\\
10855512&289.20384&48.20114&12.88&2.72&4724&0.00&65.23&12.09&0.96\\
10936814&298.47315&48.39799&10.79&2.47&4922&0.07&38.30&2.60&4.45\\
11140831&293.52471&48.79779&12.83&2.45&4862&-0.38&32.48&42.32&0.27\\
11145672&295.58421&48.79117&11.95&2.69&4915&-0.16&58.86&29.96&0.39\\
11177729&284.270741&48.87462&11.61&2.45&4896&0.07&32.84&42.99&0.27\\
11192141&292.719624&48.852017&10.52&2.01&4283&0.00&9.13&17.10&0.68\\
11287896&286.655441&49.03027&12.92&2.42&4947&0.07&34.53&55.10&0.21\\
11353223&293.29443&49.16571&12.47&2.63&4620&0.16&50.02&81.24&0.14\\
11400880&290.779849&49.27629&12.82&2.64&4755&-0.04&50.42&18.37&0.63\\
11567797&295.93539&49.51963&13.35&2.41&4816&0.07&30.69&5.60&2.07\\
11663151&292.58886&49.76146&11.96&2.68&4887&-0.24&36.30&47.11&0.25\\
11953849&285.76821&50.31719&11.20&2.45&4896&0.07&33.78&84.45&0.14\\
12003253&285.4657995&50.41648&11.19&2.41&4785&-0.24&34.27&4.48&2.58\\
12056767&288.59802&50.52476&10.14&2.45&4817&0.07&37.02&58.15&0.20\\
12067693&294.63153&50.55264&12.22&2.72&4862&0.36&29.97&65.34&0.18\\
12117920&295.27409&50.68874&12.18&2.45&4996&0.07&39.41&27.16&0.43\\
12645236&289.38545&51.76019&12.54&3.17&5114&-0.40&40.33&32.65&0.35\\
12737382&290.70515&51.90956&13.70&3.54&5023&-0.34&191.43&46.30&0.25\\
\hline
\end{tabular} 
\end{center}
\end{small}
\end{table*}

\begin{table*}
\begin{small}
\begin{center}
\caption{Details of the \confirmedblends\ confirmed chance alignments with a known entry in the KIC. In cases where we found the anomalous peak to be part of $\delta$-Scuti oscillations, the peak frequency ($\nu_{\rm peak}$) is marked with two asterisks.}
\vspace{0.1cm}
\begin{tabular}{r l l r r r r r r r r}        
\hline         
KIC & RA (deg) & DEC (deg) & Kp & log(g) & $T_{\rm eff}$ (K) & {[}Fe/H{]} & $\nu_{\rm max}$ ($\mu$Hz) & $\nu_{\rm peak}$ ($\mu$Hz) & $P_{\rm peak}$ (days) & Contam. KIC\\  
\hline
757076&291.03872&36.59813&11.68&3.58&5160&-0.10&271.50&31.86&0.36&757099\\
1026473&291.14901&36.72203&13.79&2.36&4788&-0.32&30.96&7.41&1.56&1026474\\
1872166&292.2872805&37.31746&11.63&2.82&4971&0.21&77.96&36.06&0.32&1872192\\
1872210&292.29633&37.31521&10.40&2.79&5250&0.21&78.82&17.28&0.67&1872192\\
2167774&293.07089&37.52352&9.85&2.69&4720&0.12&61.69&32.77&0.35&2167783\\
4071950&295.24143&39.11102&13.59&3.25&5004&-0.04&207.74&23.42&0.49&4071949\\
4077044&296.26499&39.18323&13.72&2.70&4832&-0.06&58.85&**167.62&0.07&4077032\\
4547321&287.067369&39.66424&13.93&3.18&4927&-0.04&245.22&40.12&0.29&4547308\\
4906950&285.9681405&40.06682&10.65&2.61&4655&0.14&52.14&18.80&0.62&4906947\\
5535029&292.2791805&40.74402&12.25&2.04&4462&-0.18&13.23&49.95&0.23&5535061\\
6048862&293.79036&41.3675&12.77&2.49&4941&0.07&34.46&86.85&0.13&6048876\\
8456004&299.5297905&44.40994&13.95&2.24&4562&-0.10&17.66&85.77&0.13&8456010\\
9045025&298.5382605&45.37306&12.47&2.66&4569&0.22&58.19&**164.00&0.07&9045002\\
9406638&292.86609&45.98426&11.36&2.49&4935&-0.50&39.83&90.95&0.13&9406652\\
9471796&294.5013705&46.06111&11.98&2.99&5014&0.21&103.73&12.53&0.92&9471797\\
9782817&296.54586&46.58502&11.84&2.61&4903&-0.32&53.07&80.03&0.14&9782831\\
9899421&295.23249&46.77037&12.30&2.32&4601&-0.22&26.00&34.74&0.33&9899414\\
10553525&298.23522&47.79192&12.81&2.80&4770&-0.26&76.44&16.24&0.71&10553491\\
\hline
\end{tabular} 
\label{tab:sample2}
\end{center}
\end{small}
\end{table*}

\begin{table*}
\begin{small}
\begin{center}
\caption{Details of the \blendsminus\ presumed chance alignments. In cases where we found the anomalous peak to be part of $\delta$-Scuti oscillations, the peak frequency ($\nu_{\rm peak}$) is marked with two asterisks.}
\vspace{0.1cm}
\begin{tabular}{r l l r r r r r r r}
\hline         
KIC & RA (deg) & DEC (deg) & log(g) & Kp & $T_{\rm eff}$ (K) & {[}Fe/H{]}  & $\nu_{\rm max}$ ($\mu$Hz) & $\nu_{\rm peak}$ ($\mu$Hz) & $P_{\rm peak}$ (days)\\ 
\hline
1870196&291.8805&37.34748&12.65&3.20&4895&0.10&191.59&60.77&0.19\\
2018906&292.42326&37.428&13.20&3.30&5046&-0.54&155.70&44.07&0.26\\
2163856&292.22694&37.55744&11.70&2.78&5010&0.07&72.14&144.29&0.08\\
2301349&291.09195&37.64004&13.43&2.72&4615&0.36&64.62&33.95&0.34\\
2569650&290.19609&37.81097&15.88&3.62&4986&0.22&187.60&74.04&0.16\\
2569935&290.222441&37.80783&13.12&1.55&4082&0.36&5.21&71.10&0.16\\
2696115&286.8690795&37.95218&11.85&2.19&4619&0.21&20.41&42.07&0.28\\
2710194&290.79134&37.92911&12.15&2.68&4580&0.24&54.74&33.60&0.34\\
\hline
\end{tabular}
\end{center}
\end{small}
\end{table*}

\begin{table*}
\begin{small}
\begin{center}
\begin{tabular}{r l l r r r r r r r}
\hline         
KIC & RA (deg) & DEC (deg) & log(g) & Kp & $T_{\rm eff}$ (K) & {[}Fe/H{]}  & $\nu_{\rm max}$ ($\mu$Hz) & $\nu_{\rm peak}$ ($\mu$Hz) & $P_{\rm peak}$ (days)\\ 
\hline
3118806&291.95328&38.21967&10.99&2.18&4569&0.36&18.05&57.58&0.20\\
3660820&295.19049&38.76184&11.87&2.40&4969&0.07&31.36&69.17&0.17\\
3858850&293.68329&38.98237&12.01&2.31&4523&0.36&20.33&83.69&0.14\\
3866844&295.5547695&38.99418&12.65&2.42&4710&0.36&33.33&64.32&0.18\\
3953330&291.242981&39.09661&12.55&2.28&4627&-0.44&16.58&6.30&1.84\\
3955590&291.86157&39.01267&10.34&2.19&4673&0.21&97.25&21.22&0.55\\
4059983&292.2911&39.10751&13.44&2.83&4842&0.56&31.96&51.18&0.23\\
4072864&295.445499&39.12097&13.81&3.34&4961&-0.06&183.22&12.17&0.95\\
4136374&284.8337805&39.21037&10.84&2.62&4876&0.07&48.53&7.32&1.58\\
4350501&287.07155&39.41622&11.74&3.06&5016&-0.22&141.40&86.65&0.13\\
4482738&296.13902&39.57963&12.95&3.06&4936&-0.44&141.34&53.44&0.22\\
4937770&295.47659&40.03605&13.15&2.90&4924&-0.32&91.11&63.60&0.18\\
4951617&298.4266905&40.06777&10.89&2.68&4822&0.04&43.31&110.44&0.10\\
5024414&295.316379&40.18652&12.71&2.81&5000&0.07&71.39&147.05&0.08\\
5112880&295.362821&40.20787&12.29&2.30&4501&0.10&27.71&66.38&0.17\\
5219666&299.3242695&40.37239&12.59&2.64&4665&0.36&56.52&29.74&0.39\\
5304555&298.761761&40.45563&12.78&2.56&4840&0.21&46.76&75.55&0.15\\
5308777&299.5773105&40.49851&13.20&2.84&4877&0.14&86.42&12.25&0.94\\
5385245&297.54492&40.55547&10.96&3.08&5083&-0.14&124.83&6.83&1.70\\
5556726&297.62054&40.76302&12.17&3.21&4898&0.07&207.26&24.04&0.48\\
5561523&298.632431&40.79503&13.47&2.99&4948&-0.06&34.49&73.03&0.16\\
5598645&283.7800305&40.81878&11.69&3.21&4973&-0.14&261.85&17.84&0.65\\
5648894&298.88994&40.82022&8.58&2.81&5068&-0.36&74.87&30.16&0.38\\
5725960&297.1535&40.92263&12.26&3.45&5171&-0.28&242.25&67.00&0.17\\
5736093&299.20563&40.91081&13.02&2.93&5121&0.07&106.57&50.05&0.23\\
6105113&284.81205&41.41103&13.24&3.10&4754&-0.02&32.89&52.65&0.22\\
6382830&296.992239&41.7899&11.01&2.24&4716&0.21&22.17&56.83&0.20\\
6447614&293.355371&41.88883&13.85&3.47&5132&-0.64&27.02&85.11&0.14\\
6612644&293.71154&42.01754&12.51&2.60&4681&0.28&44.85&18.77&0.62\\
6701238&294.48732&42.16711&11.30&2.40&4777&-0.30&27.26&97.46&0.12\\
6952355&292.99371&42.43124&13.11&2.98&4864&0.00&118.39&27.38&0.42\\
6963285&295.8375&42.49793&13.51&2.75&5102&0.21&181.09&57.55&0.20\\
7198587&291.2658405&42.73949&12.89&3.18&4968&0.18&168.98&20.42&0.57\\
7335713&280.902429&42.93572&12.44&3.13&5261&-0.56&155.32&40.35&0.29\\
7461601&296.29989&43.07131&13.43&2.90&4893&-0.46&50.17&82.46&0.14\\
7604896&290.9272605&43.23875&13.09&2.92&4878&0.06&99.38&72.13&0.16\\
7630743&297.9376695&43.28479&12.62&3.14&4786&0.07&166.91&86.71&0.13\\
7631194&298.05033&43.23317&11.82&2.68&4994&0.07&58.83&57.06&0.20\\
7831725&294.8964&43.52991&12.87&2.66&4985&0.07&39.10&35.81&0.32\\
7880664&287.7593895&43.68936&12.36&2.63&4802&0.07&39.26&99.11&0.12\\
7944142&284.869578&43.733161&7.82&2.80&5055&0.07&74.88&6.69&1.73\\
8052184&298.7172&43.85774&13.51&3.56&5196&-0.30&253.94&78.51&0.15\\
8409750&281.62728&44.41363&12.20&3.27&5245&-0.18&206.51&36.58&0.32\\
8649099&299.32878&44.79816&11.23&2.54&4845&-0.20&44.92&74.07&0.16\\
8914107&300.59514&45.15587&12.08&3.12&4910&-0.06&165.07&230.17&0.05\\
9091772&292.936899&45.41815&11.53&2.95&5026&0.07&115.53&48.25&0.24\\
9291830&295.97199&45.71479&11.02&2.56&5038&0.07&43.86&85.70&0.14\\
9479404&297.08928&46.03567&9.83&2.85&5191&0.21&77.41&10.02&1.16\\
9582089&289.24935&46.26023&12.22&2.45&4608&0.24&23.72&80.49&0.14\\
9612084&299.8694295&46.24609&11.76&3.05&5095&-0.06&78.60&**265.85&0.04\\
9771905&292.40451&46.5417&11.71&2.15&4393&0.16&11.47&**238.68&0.05\\
9906673&297.57587&46.76234&12.74&2.69&4937&-0.22&37.73&27.36&0.42\\
10203751&290.21211&47.202&11.96&2.58&4816&-0.08&35.69&13.18&0.88\\
10528911&289.28019&47.70018&12.06&2.39&4889&0.21&33.33&60.12&0.19\\
10854977&288.95117&48.22104&13.81&3.19&5193&-0.24&181.64&10.47&1.11\\
10858675&290.6682705&48.20462&12.24&2.84&4967&-0.08&87.12&24.31&0.48\\
10878851&298.14741&48.29483&13.21&2.44&4981&0.07&34.02&79.74&0.15\\
11298371&292.571829&49.03284&10.26&2.44&4989&0.07&35.28&59.76&0.19\\
11618859&295.67499&49.61104&13.77&3.64&5355&0.04&259.83&33.80&0.34\\
11753010&285.65682&49.90863&11.01&2.06&4303&0.14&14.03&89.70&0.13\\
12117138&294.895719&50.60238&12.50&2.53&4805&0.07&39.99&2.63&4.40\\
\hline
\end{tabular} 
\label{tab:sample3}
\end{center}
\end{small}
\end{table*}

\end{document}